# New quantum phases of matter: Topological Materials

Vishal Bhardwaj and Ratnamala Chatterjee

Department of Physics IIT Delhi, Hauz Khas 110016 India

**In this article, we provide an overview of the basic concepts of novel topological materials. This new class of materials developed by combining the Weyl/Dirac fermionic electron states and magnetism, provide a materials-science platform to test predictions of the laws of topological physics. Owing to their dissipationless transport, these materials hold high promises for technological applications in quantum computing and spintronics devices.**

In condensed matter physics phase transitions in materials are generally understood using Landau's theory of Symmetry breaking. For example phase transformation from gas (high symmetric) to solid(less symmetric) involves breaking of translational symmetry, another example is ferromagnetic materials that spontaneously break the spin rotation symmetry during a phase change from paramagnetic state to ferromagnetic state below its Curie temperature.

However with the discovery of Quantum Hall effect in 2D systems, it was realized that the conducting to insulating phase transitions observed in Hall conductivity ($\sigma_{xy}$) do not follow the time reversal symmetry breaking condition. Thus it was realized that a new classification of matter based on topology of materials is required to understand quantum phase transitions in quantum Hall Effect. These new quantum phases of matters that are characterized using topology are referred to as topological phases of matter. Topology[1] is the branch of pure mathematics that studies the properties of objects that are invariant under smooth deformations. A topological phase is always associated with a topological invariant which cannot change as long as there is a continuous change of parameters. For instance 2D surfaces can be topologically classified by counting their genus *g*, which counts the number of holes. Cuboid and sphere shown in Fig. 1 belong to same topological phase with g=0, both can be transformed into each other without poking a hole, however the topology of sphere (g=0) and toroid (g=1) is different, see Fig.1. Here *g* acts as topological invariant for classifying topology of 2D surfaces. Quantized value of the Hall conductance and number of gapless boundary modes obtained in quantum Hall effect are insensitive to smooth changes in material parameters and can change only when the system passes through a quantum phase transition (change in topology). The insulating phase observed in quantum Hall Effect is different from the usual band insulators which can be understood in realm of Landau's theory of symmetry breaking. Thouless, Kohmoto, Nightingale and den Nijs discussed in 1982 this difference between band insulators and these new quantum insulating states in terms of topology[2]. The phase transformation between different quantum hall states does not break any symmetry but can be defined using topology change using an integer called TKNN invariant *ν*. The *ν* provides robust quantization to Hall conductivity ($\sigma_{xy}$) measured in quantum

Hall effect as per relation $\sigma_{xy}=\nu(e^2/h)$. A point to be noted here is that the Quantum Hall States can be realized only in presence of external magnetic fields at very low temperature.

The quest among researchers to obtain these new topologically protected quantum states even in the absence of external magnetic field gave rise to the discovery of Topological insulators[3]. Topological insulators are solid-state materials that are insulators in the bulk but have intrinsic surface states that behave like metal, and are completely robust to any type of defects or disorder. The role of external magnetic field is played by a fictitious magnetic field induced by large spin orbit coupling of one of the elements in this material system[4]. The spin orbit interactions inside a material bring bulk band inversion around Fermi level and provide non-trivial topology to the bulk bands. The non-trivial topology of bulk bands result in evolution of time reversal symmetry protected surface states near the system boundary.

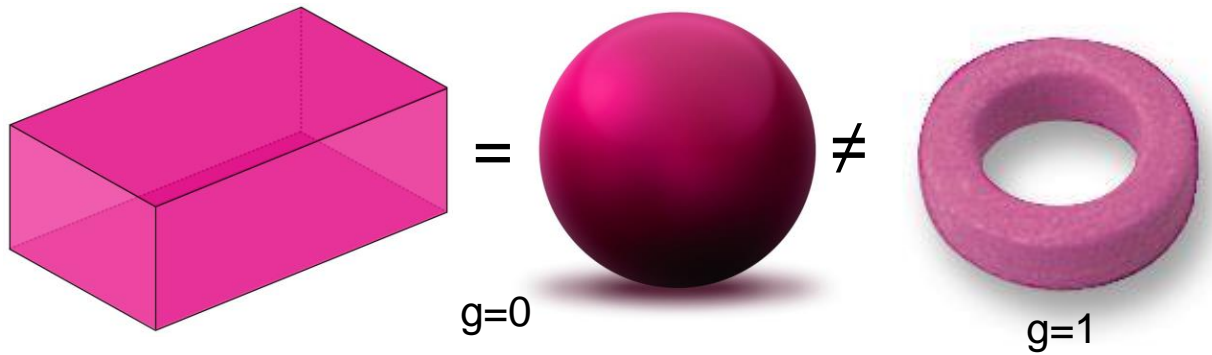

**Fig.1** Topology of cuboid and sphere is same with number of holes g=0, whereas topology of toroid is different with g=1.

The topological order parameter used to characterize Topological phases of matter is Berry's phase. The idea of the Berry's phase was first introduced in the context of adiabatic approximation, where the system is subjected to a very slowly varying perturbation in time. The Berry's phase is defined mathematically as the line integral of Berry curvature of valence band Bloch wave functions integrated over the first Brillouin zone. The TKNN invariant or the first chern number used to define Quantum Hall States is also related to Berry's phase and is equal to Berry phase divided by $2\pi$. The significance of Berry's phase for classification of topological insulator and topological semimetals is discussed in the next sections of this article. Let's first discuss the topological insulators.

## Topological Insulators:

Topological insulators are a new quantum phase of matter with insulating bulk and conducting surface/edge states in three/two dimensions, see Fig.2(a-b). These surface states are very different from an ordinary metallic state where up and down spins are distributed everywhere on a Fermi surface. The surface states in topological insulator have spin degeneracy and there are separate channels for up and down spins with their momenta locked to spins perpendicularly to preserve time reversal symmetry, see Fig. 2(a-b). The surface/edge states are robust against the non-magnetic impurities due to the topological protection, which result in dissipation less and back scattering prohibited high mobility carrier transport. Evolution of metallic surface or edge states at the system boundary of topological insulators can be explained using bulk boundary correspondence process. When the gapped energy states of bulk having non-trivial topology are terminated with a trivial insulator (*e.g* Vacuum), the topological invariant changes at the interface. In this process of the topology change from non-trivial to trivial, the energy gap closes at the interface and the metallic surface states appear. Hence, in 3D/2D Topological insulators the gapless (metallic) surface/edge states are always present.

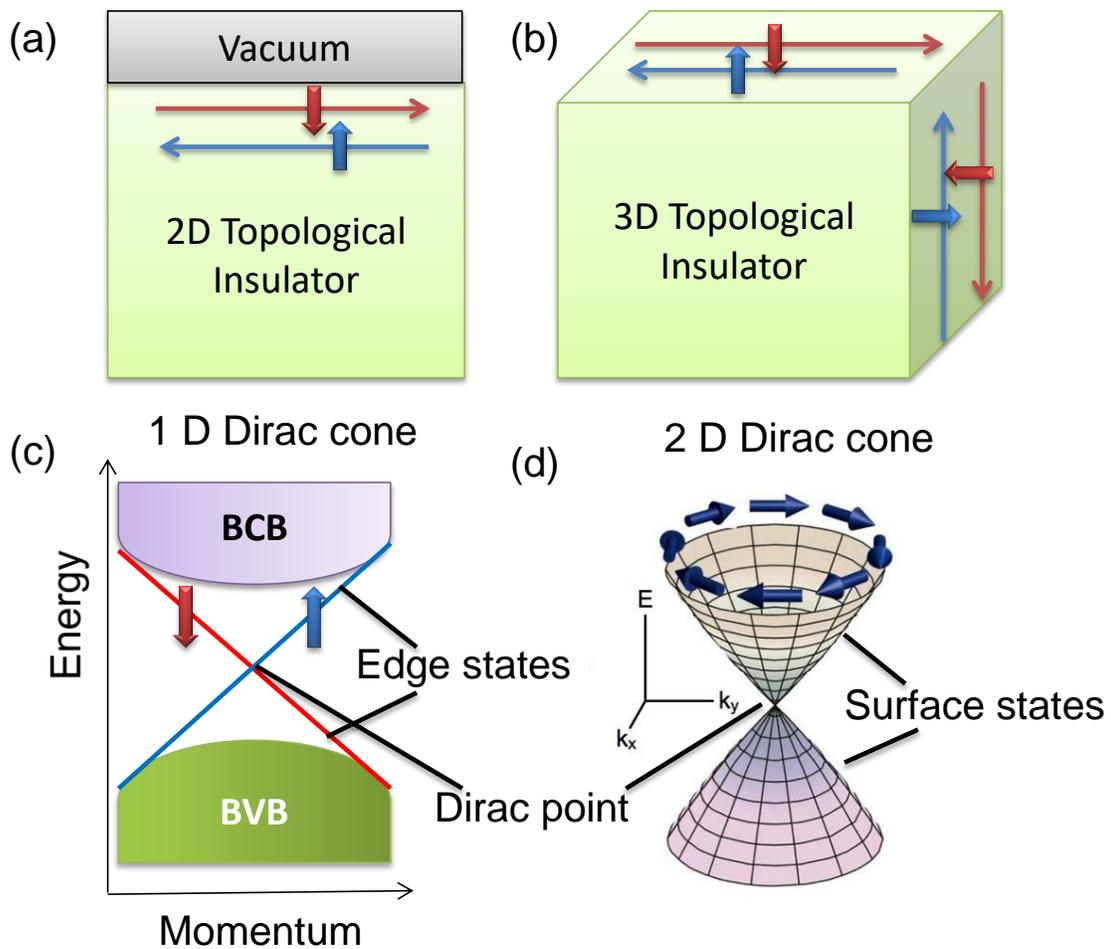

**Fig.2**(a-b) Schematic real space diagram of 2D and 3D topological insulators with spin polarized edge and surface states at the system boundary. (c-d) Energy band diagram of 2D and 3D topological insulator in momentum space, showing the formation of 1D and 2D Dirac cone, respectively. BCB: Bulk conduction band and BVB : bulk valence band.

It was realized that the non-trivial topology of electronic states for topological insulators can be fully characterized using one $Z_2$ invariant in 2D system and four $Z_2$ invariants in 3D system[5]. The $Z_2$ invariant gives topological classification based on parity which can be understood easily using Kramer's theorem for spin ½ electrons. For time reversal symmetry protected systems the energy at +k and –k is the same *i.e* the energy bands come in pairs which are called Kramer's pairs. These pairs are degenerate at certain points in Brillouin zone that are referred to as time reversal invariant momentum points or Kramers degenerate points, due to the periodicity of Brillouin zones. By counting how often the surface states cross the Fermi energy between two boundary Kramers degenerate points, one may distinguish topological nontrivial (odd number of crossings; $Z_2$=1) and trivial (even number of crossings; $Z_2$=0) system. In the nontrivial case, edge or surface states crosses each other odd number of times and this point of crossing is called the Dirac point, see Fig. 2(c-d). The spin degeneracy is lifted around this point which results in formation of 1D and 2D Dirac cone due to the edge and surface states in 2D and 3D topological insulators respectively see Fig.2(c-d). As the electron traverse a circular path around the Dirac point it acquires a nontrivial Berry's phase=π. Hence, the experimental realization of Berry's phase=π indicates the presence of Dirac cone and Dirac fermions in topological materials.

**Topological Semi-metals:**

Semi-metals have band structure in which the conduction and valence band touch each other at a point but the carrier concentration is very less ($10^{17}$-$10^{22}$cm$^{-3}$) in comparison to metals ($10^{22}$-$10^{25}$cm$^{-3}$). Some of these semi-metals have a narrow band gap between the highest occupied and the lowest unoccupied band along with band touching points (nodes) at the Fermi energy[6] in the first Brillouin zone. There can also be Line nodes, where the bands are degenerate along closed lines in momentum space in the semimetals. Now if these points or line nodes also have topological protection due to nontrivial bulk band structure, then the semimetals are classified as topological semimetals. Topological semi-metals are the quantum phases of matter that host Dirac and Weyl fermions. These are further classified into Weyl semi-metals, Dirac semi-metals, triple point semi-metals and nodal line semi-metals[6]. The weyl semi-metals are currently the most studied topological semi-metals with Weyl fermions as the charge carriers. The non-centrosymmetric semi-metals like TaAs(inversion symmetry-breaking) or magnetic semi-metals like Heusler alloys (time-reversal symmetry breaking) are the potential candidates for the quest of the Weyl semi-metals. These semi-metals have two singly degenerate bulk band crossings called as Weyl nodes at particular values of crystal momentum in the first Brillouin zone. These band crossings disperse linearly in all directions of momentum space away from the Weyl nodes. Weyl nodes act as monopole (source) and anti-monopole (sink) of the Berry curvature field in the momentum space see Fig. 3. They are always separated in momentum space and appear in pairs of positive and negative chirality. Similar to topological insulator the boundary of Weyl semimetals have gapless surface states, which are topologically protected by chiral charge associated with the Weyl nodes present in bulk bands. This is also an example of the bulk-boundary correspondence in this topological phase. These surface states are called Fermi arcs in Weyl semi-metals which connect the pairs of bulk Weyl nodes with opposite chiralities and provide a surface fingerprint of the topological nature of the bulk band structure, see Fig. 3.

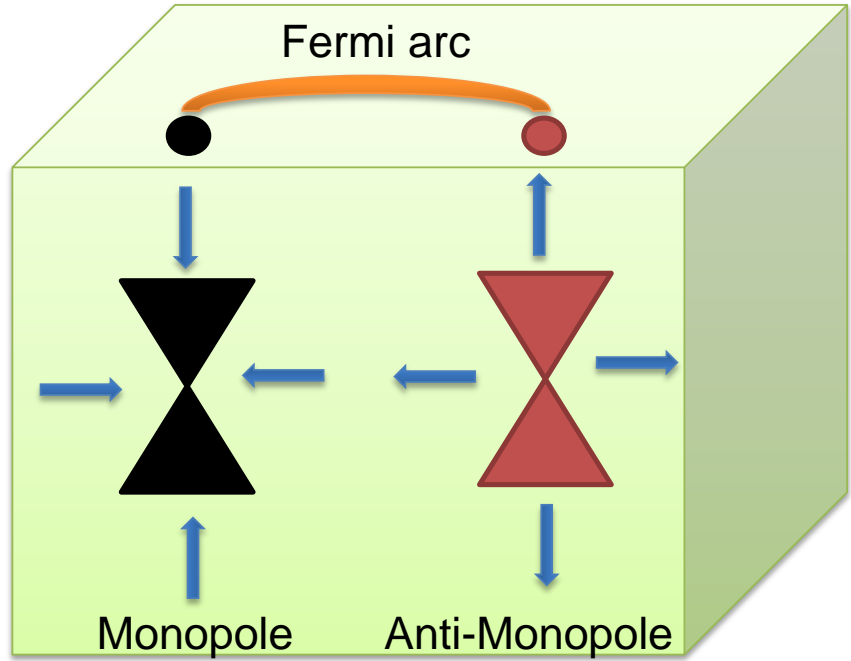

**Fig. 3** Schematic diagram of a Weyl semimetal which have two Weyl nodes in bulk bands.

The chiral charge associated with a Weyl node is the topological invariant used to define topology of Weyl semi-metals. Weyl semi-metals do not have a global band gap due to the existence of two Weyl nodes; therefore, the topological invariant can't be defined for the entire 3D bulk Brillouin zone. Hence, it is defined on a 2D closed surface that encloses one Weyl node in momentum space where the band structure is fully gapped. The chiral charge is calculated mathematically by the surface integral of the Berry curvature associated with the Weyl node. A conserved chiral current of fermions with positive and negative chirality always transfers between a Weyl node pair. In an external magnetic field applied parallel to the external electric field, population imbalance of chiral fermions is created which generates a chiral chemical potential. This effect is called chiral anomaly or chiral magnetic effect and it affects the transport and thermoelectric response in Weyl semimetals[5]. Therefore experimentally chiral anomaly can be realized as negative longitudinal magneto-resistance and large thermo power. However, in order to have significant effect on observable properties the Weyl nodes must occur at or very close to the Fermi energy.

**Experimental characterization:**

The most direct evidences of topological phases are obtained through angle resolved photoemission spectroscopy (ARPES) experiment that use soft x-rays to study the band structure of the surfaces of solids. This technique is mainly suited for probing surface states as the photons knock out valence electrons present with in 1nm of sample surface[8]. The presence of Dirac cone with topologically protected surface states in Topological insulators can be verified directly using ARPES. The nontrivial topology of the surface states can also be verified by counting the surface states crossings of the Fermi energy using Kramer's theorem as discussed above. Direct observation of the non-degeneracy of Dirac cone and helical spin polarization of the surface states can be done using spin-resolved ARPES. In case of Weyl semi-metals the soft x-ray ARPES (reasonably bulk sensitive) and vacuum ultraviolet ARPES can be used to probe Weyl nodes in the bulk and Fermi arcs on the surfaces, respectively[9]. The ultraviolet ARPES is extremely surface sensitive technique which uses low photons energy to knock out valence electrons.

Samples for ARPES experiments require clean and flat surfaces which are usually obtained by cleaving single crystals. So in cases when the materials do not cleave well or single crystals are not available the magneto-transport experiments provide an indirect method to characterize the non-trivial nature of surface states in Topological insulators and weyl-semimetals. The Berry's phase=$\pi$ is associated with massless Dirac fermions of surface states in topological insulators which provide them non-trivial nature. Now experimental verification of $\pi$ Berry's phase can be done by measuring magneto-resistance in perpendicular magnetic field(see Fig. 4(a)), where two important signatures can be looked into. First is the observation of weak antilocalization effect around low magnetic field due to $\pi$ Berry's phase associated with the charge carriers, see Figs. 4(a) and 4(b). Second is the presence of Shubinkov-de Hass (SdH) oscillations at high magnetic fields due to landau level quantization, see Figs. 4(a) and 4(c). The phase factor of these oscillations directly reflects the Berry's phase of non-trivial surface states. In weak-antilocalization effect the magneto-conductance data shows peak around zero field due to quantum interference of coherent transport wave functions of carriers, as shown in Fig. 4(b). The magnetic-field dependence of the experimental data can be described by Hikami-Larkin-Nagaoka theory, see black line fit in Fig. 4(b). The parameters($\alpha$ and $l_\phi$) obtained after fitting give information about dimensionality(2D or 3D) of topologically insulator along with nontrivial nature of surface states[10]. In presence of large magnetic field (H) the motion of charge carriers gets quantized which result in Landau level quantization. Hence quantum oscillations called as SdH oscillations are observed which are periodic as a function of 1/H, see Fig. 3(c). The Berry's phase can be estimated by fitting the experimental data to standard Lifshitz Kosevich equation or by using Landau's fan diagram analysis[10], see Fig. 1(d). However, since magneto-transport probes the interior of the sample one should verify the non-trivial nature of surface states through above both methods.

The topological semimetals have chiral anomaly associated with the Weyl nodes which provide the experimental signature in magneto-transport experiments. These signatures[12] are (i) large negative longitudinal magneto-resistance[11], (ii) Aniostropic magneto-resistance narrowing and (iii) Planar hall effect. Since, Weyl nodes act as source and sink of Berry flux this will give rise to intrinsic hall component in magnetic Weyl semimetal. Hence, observation of anomalous Hall Effect serves as the experimental tool in studying the topology of Weyl nodes in magnetic Weyl semimetals[12]. SdH oscillations in magneto-resistance and thermo-power provide a method to estimate mobility, effective mass of the carriers and Berry's phase associated with them.

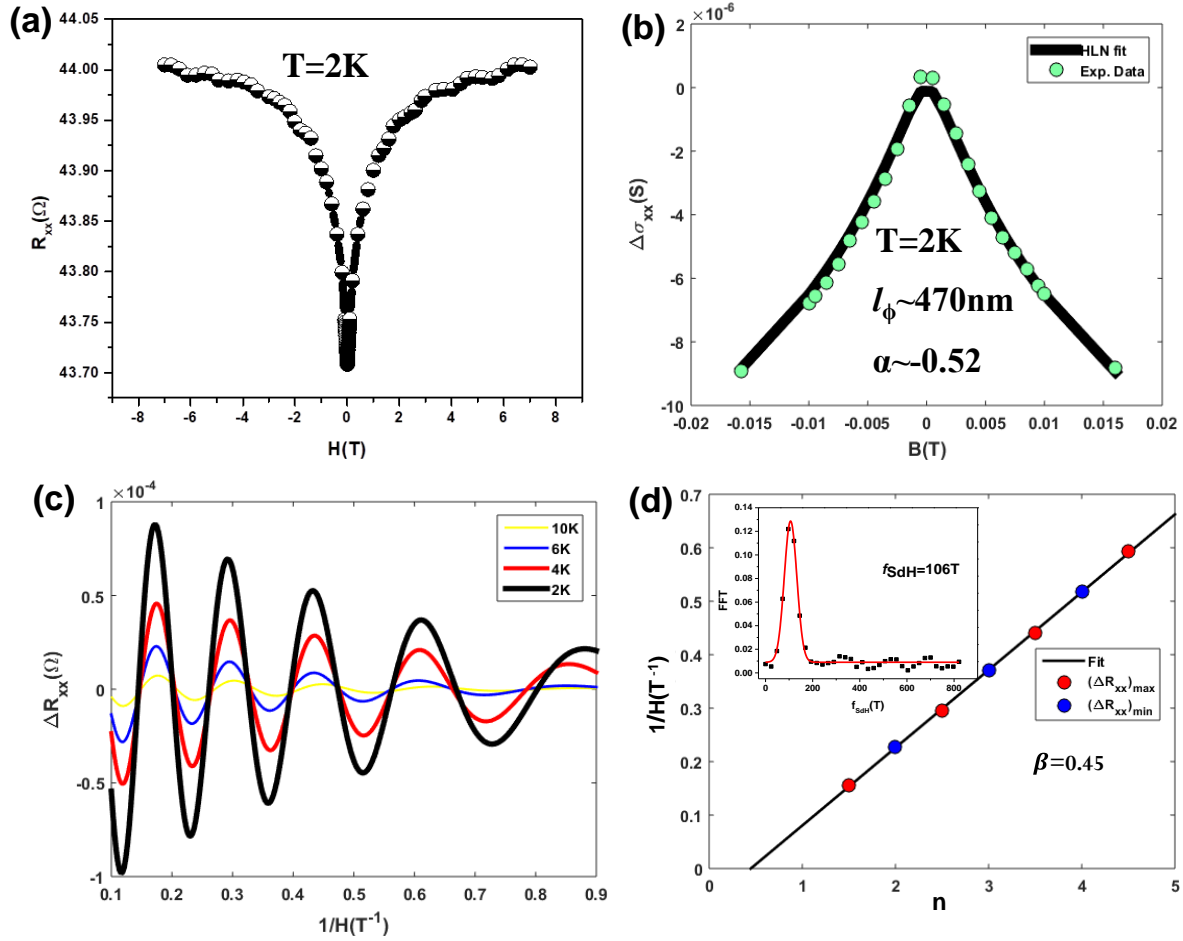

**Fig. 4** (a) Magneto-resistance data of DyPdBi thin films measured at 2K shows the observation of SdH oscillations at high magnetic fields and weak-antilocalization effect at low magnetic fields.(b) WAL effect observed in magentoconductance data. Experimental data is fitted to HLN model.(c) SdH oscillations as a periodic function of 1/H at different temperatures for DyPdBi thin films.(d) Landau-level fan diagram analysis of SdH oscillations measured at 2K to estimate Berry's phase, inset shows frequency of oscillations estimated using fast Fourier transformation. Figures are adapted from ref. 10.

## Looking forward:

In last decade we have discovered many topological phases in condensed matter physics but this could be a tip of the iceberg. The discovery of new topological phases requires band structure calculation using Density functional theory and we believe in future many new topological phases/materials would be discovered. From application point of view topological insulators and topological semimetals hold promise for dissipation-less transport through spin polarized channels. These materials have great potential in future spintronics devices and quantum computing applications. Since most of these topological phases exist at low temperatures, the aim is to push their operation toward room temperature. This would require a collective effort in experimental, theoretical and computation research field. Discovery and experimental realization of new topological phases/materials for spintronics devices will move us toward achieving this goal.

## Suggested Reading: